\documentclass[11pt]{article}
\usepackage{graphicx,citesort,ils2008}
\usepackage{amssymb}
\usepackage{amsmath}
\begin{document}
\hfill{RUB-TPII-02/08}

\begin{center}

{\LARGE\textbf{Analyticity of QCD observables beyond leading-order
               perturbation theory\footnote{Invited contribution in
               memory of I.L.\ Solovtsov at Seminar in Bogoliubov 
               Laboratory of Theoretical Physics, JINR, 141980 Dubna,
               Russia, Jan. 17, 2008.}}}
\vskip7.5mm

{\large N.G. Stefanis}

\vskip5mm

\textit{Institut f\"{u}r Theoretische Physik II,\\
        Ruhr-Universit\"{a}t Bochum,
        D-44780 Bochum, Germany\\
        Electronic address: stefanis@tp2.ruhr-uni-bochum.de}

\vskip7.5mm

{\large A.I.~Karanikas}

\vskip5mm

\textit{University of Athens, Department of Physics,\\
        Nuclear and Particle Physics Section,
        Panepistimiopolis,
        GR-15771 Athens, Greece\\
        Electronic address:akaran@phys.uoa.gr}
\vskip5mm

\textit{University of Athens, Department of Physics,\\
        Nuclear and Particle Physics Section,
        Panepistimiopolis, GR-15771 Athens, Greece}

\vskip10mm

\textbf{Abstract}

\vskip2.5mm

\parbox[t]{110mm}{{\small
A theoretical framework is presented to treat hadronic observables
within analytic perturbative QCD beyond the leading order of the 
coupling and for more than one single large momentum scale.
The approach generalizes and extends the pioneering work of
Shirkov and Solovtsov on an analytic strong running coupling.
Some applications to hadronic observables at the partonic level
are also discussed.}}

\end{center}

\vskip5mm
\section{Homage to Igor Solovtsov}
\label{Sect:memorial}
I (NGS) met Igor for the first time many years ago in the beginning
of the nineties during a physics conference in Dubna.
It was a chance encounter during the conference dinner, when it
happened to get a seat next to him.
Igor introduced himself and I took the opportunity to discuss with him
about variational perturbation theory and its convergence properties,
a subject on what he was working at that time, known to me from his
publications.
We soon found out that we had a lot of common interests in physics
and we kept discussing for hours, while emptying a bottle of vodka.
During this first encounter, I could, of course, not imagine that years
later I would engross myself so strongly in Analytic Perturbation
Theory, a subject pioneered by Igor and Dmitry V.\ Shirkov, as it
actually happened.
This activity will be surveyed in this invited contribution to his
memory.
Since then, I met Igor on several occasions, in Dubna and abroad,
and I still have a strong recollection of our discussions.
Igor indelibly forged his name in the annals of physics by his
scientific achievements---no doubt.
But those of us, who had the privilege to know him in person, will 
sadly remember and miss his intellectual creativity along with his 
kindness.

\section{Introduction}
\label{Sect:intro}

Traditionally, perturbation theory in QCD suffers from an artificial
singularity at momenta close to $\Lambda_{\rm QCD}$, called (at the
one-loop level) the Landau pole.
Since the early days of QCD, theorists have tried different remedies
to avert this problem, like infrared (IR) cutoffs, ``freezing'', etc.
But approximately ten years ago, Shirkov and Solovtsov have devised 
an approach that avoids this problem by appealing only to a few
basic principles of Quantum Field Theory---chief among them, Causality
and Renormalizability---and avoiding the introduction of extraneous
infrared (IR) regulators.
Since this pioneering work appeared in the year 1996 \cite{SS96,SS97},
the analytic approach to QCD perturbation theory has evolved and
considerably progressed, shifting the cutting edge significantly (see 
\cite{SS99,Shi01,SS06} for reviews and further references), and finally 
culminating into the so-called Analytic Perturbation Theory (APT).
Meanwhile this analytic approach has been extended beyond the one-loop
level \cite{MiSol97,SS98} and important techniques for numerical
calculations have been developed
\cite{Mag99,Mag00,KM01,Mag03u,KM03,Mag05}.
Also applications to the ultra-low momentum region have been carried
out, e.g., \cite{BNPSS07}, and alternative formulations of the strong
coupling below the Landau pole have been proposed aiming to
incorporate nonperturbative input \cite{Nes03,NP04}.

The simple analytization concept of the strong running coupling has
been generalized to the level of hadronic amplitudes \cite{KS01,Ste03}
and new techniques have been developed to deal with more than one
large hard scale in the process \cite{SSK99,SSK00}, including also
Sudakov resummation in exclusive processes.
In the later course of these investigations, it was realized that
logarithms of the aforementioned second large scale---which can be the
factorization or the evolution scale---correspond to non-integer
(fractional) powers of the coupling, giving rise to Fractional
Analytic Perturbation Theory (or FAPT for short) \cite{BMS05,BKS05}.
At the heart of this development was the Karanikas-Stefanis (KS)
\cite{KS01} analytization principle which demands that all terms in a
QCD amplitude that can affect the discontinuity across the cut along
the negative real axis $-\infty<Q^2<0$, and hence contribute to the
spectral density, have to be included into the analytization
procedure, i.e., the dispersion relation.
The KS procedure encompasses the Shirkov-Solovtsov analytization concept
of integer powers of the strong coupling and paves the way to the
analytization of any real power both in the Euclidean \cite{BMS05} as
well as and in the Minkowski region \cite{BMS06}.
The crucial advantages of this scheme are:
\begin{itemize}
\item A diminished sensitivity on the factorization scale of typical
      QCD hard processes, like the factorized part of the pion's
      electromagnetic form factor---verified in \cite{BKS05} to
      the next-to-leading order (NLO).
\item A quasi renormalization-scale setting and scheme independence of
      the same observable at the NLO level \cite{BPSS04} (see for an
      abridged version \cite{Ste04mon}).
\item The inclusion of evolution effects due to the running of the
      strong coupling beyond the one-loop order \cite{BMS06}.
\item A faster convergence of the perturbative expansion in terms of
      analytic images of real powers of the strong coupling
      \cite{BMS05}.
\item The resummation to all orders of the $\pi^2$ terms, induced by
      analytic continuation into the timelike domain.
      This has been exemplarily verified in \cite{BMS06} for the scalar
      Higgs decay into a $b\bar{b}$ pair at the four-loop level.
\end{itemize}

The most important consequence of the analytic approach is that it
ties the abstract mathematical requirements of causality and
renormalizability to something tangible, like the calculation of the
Bjorken \cite{MSS98} and the Gross-Llewellyn Smith sum rule
\cite{MSS98GLS}, or the inclusive decay of a $\tau$-lepton into
hadrons \cite{JS95-349,MSS97,MSSY00}, the pion's electromagnetic form
factor \cite{SSK99,SSK00,BPSS04,BKS05}, the Higgs boson decay into a
$b\bar{b}$ pair \cite{BMS06}, and many other processes.
In the present exposition we will present the bedrock of this approach,
focusing our attention to selected applications and results beyond the
leading order (LO) of perturbative QCD---conventional and analytic.

\section{Basic Structure of FAPT}
\label{Sect:FAPT}

Let us briefly review FAPT, specifying our notation, explaining its
main principles, and describing its methodology.

The running strong coupling in QCD
\begin{equation}
  \alpha_s(Q^2)
=
  \frac{4\pi}{\beta_0}a_s(L)
  ~~~{\rm with}~~~
  L
\equiv
  \ln \frac{Q^2}{\Lambda^2} \, ,
\label{eq:alpha}
\end{equation}
%Eq (1)
where $\Lambda$ denotes the characteristic scale of QCD,
$\Lambda_{\rm QCD}$,
satisfies the renormalization-group (RG) equation
\begin{eqnarray}
 \frac{d}{dL}\left(\frac{\alpha_{s}}{4 \pi}\right)
=
  \beta\left( \frac{\alpha_{s}}{4 \pi}\right)
= - b_0\left(\frac{\alpha_{s}}{4 \pi}\right)^2
  - b_1\left(\frac{\alpha_{s}}{4 \pi}\right)^3
  - b_2\left(\frac{\alpha_{s}}{4 \pi}\right)^4\,- \ldots 
\label{eq:betaf}
\end{eqnarray}
%Eq (2)
with known $\beta$-function coefficients up to the displayed order.
[Their explicit expressions can be found, for instance, in
\cite{BMS06}].
At the one-loop order, $a_s(L)$ develops a Landau pole, while the
two-loop solution of Eq.\ (\ref{eq:betaf}) has a square-root
singularity, with more complicated singularities for still higher
orders.
Shirkov and Solovtsov \cite{SS97} have shown that the ghost-singularity
problem can be solved only on account of renormalizability---in terms
of the RG equation---and causality---expressed in the form of a
dispersion relation.
Then, one obtains in Euclidean space analytic images of the coupling
at loop order $l$
\begin{eqnarray}
  {\cal A}^{(l)}_{m}(Q^2)
\equiv
   \left[a_{(l)}^m(Q^2)\right]_{\rm an}
\label{eq:a_caligraphic}
\end{eqnarray}
%Eq (3)
following from the K\"{a}ll\'en-Lehmann spectral representation
\begin{eqnarray}
 \left[f(Q^2)\right]_\text{an}
  =
  \frac{1}{\pi}
   \int_0^{\infty}\!
    \frac{{\bf Im}\,\big[f(-\sigma)\big]}
         {\sigma+Q^2-i\epsilon}\,
     d\sigma\
\label{eq:dispersion}
\end{eqnarray}
%%Eq (4)
with the spectral density at one loop given by (see \cite{BMS05,BMS06} 
for higher loops)
\begin{eqnarray}
  \rho_{\nu}(\sigma)
=
  \frac{1}{\pi}\,
  {\bf Im}\,\big[a^{\nu}(-\sigma)\big]
= \frac{1}{\pi} 
  \frac{\sin(\nu\phi)}{\left[\pi^2+L^2(\sigma)\right]^{\nu/2}}
  \, .
\label{eq:spec-dens-n}
\end{eqnarray}
%Eq (5)
As a result, one then finds in the Euclidean space at one loop
\begin{eqnarray}
  {\cal A}^{(1)}_{1}(Q^2)
& = &
  \frac1{L}- \frac1{e^L -1}\, ,
\label{eq:A_1}
\end{eqnarray}
%Eq (6)
while its counterpart in Minkowski space reads
\begin{eqnarray}
{\mathfrak A}^{(1)}_{1}(s)
  &=& \frac{1}{\pi}\,
       \arccos\left(\frac{L_{s}}{\sqrt{L_{s}^2+\pi^2}}\right)
\label{eq:U_1}
\end{eqnarray}
%Eq(7)
with
$
  L_s = \ln\left(s/\Lambda^2\right)
$.
This procedure is sufficient in considering QCD processes with one
large scale, but fails for non-integer powers of $a_s$ and is
unable to accommodate terms, like
\begin{equation*}
\begin{array}{lll}
  & \bullet~~~\left[a_s(L)\right]^{\gamma_0/2\beta_0} 
  & \ \longleftrightarrow  \, \, \, \, \, {\rm RG~ at~ one~ loop} \\
  & \bullet~~~\left[a_s(L)\right]^{n}\ln[a_s(L)]      
  & \ \longleftrightarrow  \, \, \, \, \, {\rm RG~ at~ two~ loops}\\
  & \bullet~~~\left[a_s(L)\right]^{n}L^m              
  & \ \longleftrightarrow  \, \, \, \, \, {\rm Factorization}     \\
  & \bullet~~~\exp\left[-a_s(L) F(x)\right]           
  & \ \longleftrightarrow  \, \, \, \, \,
      {\rm Sudakov~ resummation~ (symbolically)}
  \
\end{array}
\end{equation*}
which typically appear in perturbative calculations beyond the 
leading order due to the reasons already explained.
Such terms do not modify the ghost singularities but they do
contribute to the spectral density and, hence, their analytic images
are inevitably required for the dispersion relation.
In actual fact, precisely such terms are tantamount to
{\it fractional} (real) powers of the strong coupling
\cite{BMS05,BKS05}.

\begin{figure}[ht]
 \centerline{\includegraphics[width=0.3\textwidth]{%fig-APT.eps
  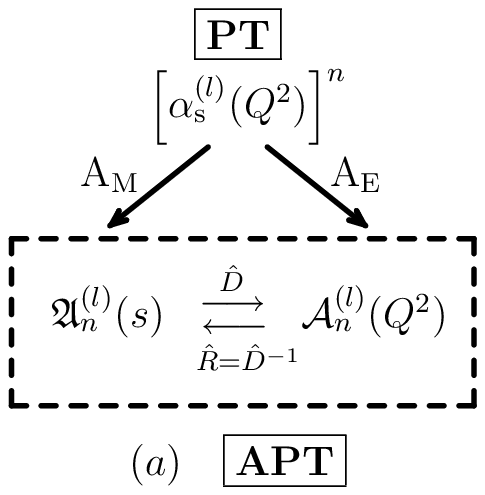}}
  \vspace*{2mm}
 \centerline{\includegraphics[width=0.3\textwidth]{%fig-FAPT.eps
 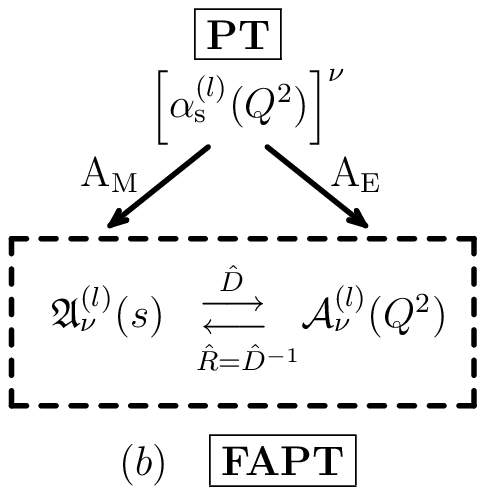}
 ~~~\includegraphics[width=0.3\textwidth]{%fig-FAPT2.eps
 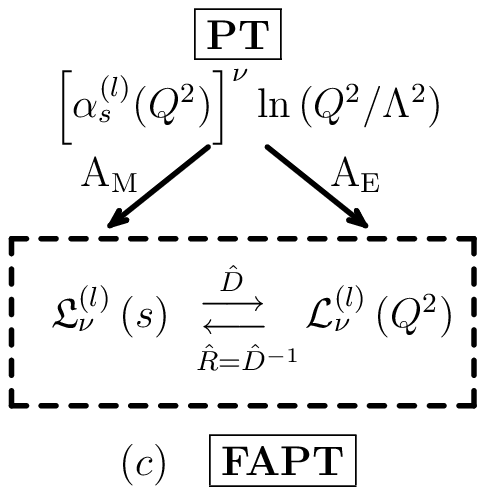}}
   \vspace{0.0cm}
   \caption{Illustration of different analytization concepts.
           (a) APT (b) FAPT, only strong-coupling powers,
           (c) FAPT, products of strong-coupling powers and
               logarithms.
   In APT, the index $n$ is restricted to integer values only; in
   FAPT $\nu$ can assume any real value. Further explanations are
   given in \protect\cite{BMS06}.
\label{fig:APT-FAPT}}
\end{figure}

The core feature of FAPT is, as mentioned in the Introduction, the KS
analytization principle \cite{KS01}.
The use of this principle allows the inclusion into the dispersion
relations of logarithmic terms, of the sort $\ln (Q^2/\mu_{\rm F}^2)$,
or products of such logarithms with powers of the running coupling.
To appreciate its meaning and usefulness, we present and compare
different analytization concepts in Fig.\ \ref{fig:APT-FAPT}.

%%%%%%%%%%%%%%%%%%%%%%%%%%%%%%FIGURE 1%%%%%%%%%%%%%%%%%%%%%%%%%%%%%%%%%
%%%%%%%%%%%%%%%%%%%%%%%%%%%%%%%%%%%%%%%%%%%%%%%%%%%%%%%%%%%%%%%%%%%%%%%

In this figure, the linear operations
$\textbf{A}_{\rm E}$ and ${\bf A}_{\rm M}$
define, respectively, the analytic running couplings in the Euclidean
(spacelike)
\begin{eqnarray}
 {\bf A}_{\rm E}\left[a^{n}_{(l)}\right]
  &=& {\cal A}^{(l)}_{n}
    ~~~{\rm with}~~~
   {\cal A}^{(l)}_{n}(Q^2)
  \equiv
   \int_0^{\infty}\!
    \frac{\rho^{(l)}_n(\sigma)}
         {\sigma+Q^2}\,
       d\sigma
\label{eq:A.E}
\end{eqnarray}
%Eq (8)
and the Minkowski (timelike) region
\begin{eqnarray}
 {\bf A}_{\rm M}\left[a^{n}_{(l)}\right]
  = {\mathfrak A}^{(l)}_{n}
  \text{~~~with~~~}
  {\mathfrak A}^{(l)}_{n}(s)
  \equiv
    \int_s^{\infty}\!
     \frac{\rho^{(l)}_n(\sigma)}
          {\sigma}\,
      d\sigma\, .
\label{eq:A.M.rho}
\end{eqnarray}
%Eq (9)

The above analytization operations can be represented by the
following two integral transformations from the timelike region
to the spacelike region (see, e.g., \cite{Shi01}):
\begin{eqnarray}
  \hat{D}\big[{\mathfrak A}^{(l)}_{n}\big]
  &=&  {\cal A}^{(l)}_{n}
    ~~~{\rm with}~~~
   {\cal A}^{(l)}_{n}(Q^2)
  \equiv
   Q^2 \int_0^{\infty}\!
     \frac{{\mathfrak A}^{(l)}_{n}(\sigma)}
          {\big(\sigma+Q^2\big)^2}\,
      d\sigma
\label{eq:D-operation}
\end{eqnarray}
%Eq (10)
and for the inverse transformation
\begin{eqnarray}
  \hat{R}\big[{\cal A}^{(l)}_{n}\big]
  &=&
  {\mathfrak A}^{(l)}_{n}
  ~~~{\rm with}~~~
  {\mathfrak A}^{(l)}_{n}(s)
  \equiv
   \frac{1}{2\pi i}
   \int_{-s-i\varepsilon}^{-s+i\varepsilon}\!
    \frac{{\cal A}^{(l)}_{n}(\sigma)}
          {\sigma}\,
      d\sigma\, .
\label{eq:R-operation}
\end{eqnarray}
%Eq (11)
These two integral transformations are connected to each other by
the relation
\begin{eqnarray}
 \label{eq:reciproc}
 \hat{D}\hat{R} = \hat{R}\hat{D} = 1\,,
\end{eqnarray}
%Eq (12)
valid for the whole set of analytic images of the powers of the
coupling in the Euclidean as well as in the Minkowski space,
$\big\{{\cal A}_n,{\mathfrak A}_n\big\}$, respectively, and at any
desired loop order of the perturbative expansion.

In the spacelike region, the analytic images of the coupling can
be expressed in terms of the reduced transcendental Lerch function
$F(z,\nu)$ to read \cite{BMS05} 
($L\equiv \ln(Q^2/\Lambda^2)$)\footnote{Everywhere in this 
presentation Greek labels denote non-integer (real) powers or 
indices.}
\begin{equation}
  {\cal A}_{\nu}(L)
      = \frac{1}{L^\nu}
      - \frac{F(e^{-L},1-\nu)}{\Gamma(\nu)}\, ,
\label{eq:lerch}
\end{equation}
%Eq (13)
where the first term corresponds to the conventional term of perturbative
QCD and the second one is entailed by the pole remover (cf.\
$1/({\rm e}^{L}-1)$ at one loop).
This function is an entire function in the index $\nu$ and has the
properties 
${\cal A}_{0}(L)=1$,
${\cal A}_{-m}(L)=L^{m}$ for $m\in\mathbb{N}$, and
${\cal A}_{m}(\pm\infty)=0$ for $m\geq 2$, $m\in\mathbb{N}$, while
for $|L|<2\pi$, it reads
$
 {\cal A }_{\nu}(L)
=
 -\left[1/\Gamma(\nu)\right]\sum_{r=0}^{\infty}\zeta(1-\nu-r)
                            \left[(-L)^{r}/r!\right].
$
In the timelike region, these images are completely determined
by elementary functions \cite{BMS05} 
($L_s\equiv \ln(s/\Lambda^2)$):
\begin{equation}
  {\mathfrak A}_{\nu}(L_s)
    = \frac{\sin\left[(\nu -1)
            \arccos\left(L_s/\sqrt{\pi^2+L_s^2}\right)
                \right]}
      {\pi(\nu -1) \left(\pi^2+L_s^2\right)^{(\nu-1)/2}}
\label{eq:gothic}
\end{equation}
%Eq (14)
from which, for example, we get
${\mathfrak A}_{0}(L_s)=1$ and ${\mathfrak A}_{-1}(L_s)=L_s$.
The salient characteristics of FAPT in comparison with APT and the
standard perturbative expansion in QCD are compiled in Table
\ref{Tab:char}, while for further reading and graphic illustrations
we refer the reader to \cite{BMS05,BKS05,BMS06}.

%%%%%%%%%%%%%%%%%%%%%%%%%%%%%%%%%%%%%%%%%%%%%%%%%%%%%%%%%%%%%%%%%%%%%%%
%                                TABLE 1                              %
%%%%%%%%%%%%%%%%%%%%%%%%%%%%%%%%%%%%%%%%%%%%%%%%%%%%%%%%%%%%%%%%%%%%%%%
 \begin{table}[h]
 \caption{FAPT versus APT and standard QCD perturbation theory (SPT)}
  \begin{tabular}{|c|c|c|c|c|}\hline
  ~~~~~~Theory~~~~~~
           & ~~~~~~~SPT~~~~~~~
                     & ~~~~~APT~~~~~
                               & ~~~~~FAPT~~~~~         \\
  \hline\hline
  Space  & $\big\{a^\nu\big\}_{\nu\in\mathbb{R}}%
           \vphantom{^{\big|}_{\big|}}$
         & $\big\{{\cal A}_m\big\}_{m\in\mathbb{N}}$
         & $\big\{{\cal A}_\nu\big\}_{\nu\in\mathbb{R}}$ 
                                                        \\
  Series expansion
         & $\sum\limits_{m}f_m\,a^m(L)\vphantom{^{\big|}_{\big|}}$
         & $\sum\limits_{m}f_m\,{\cal A}_m(L)$
         & $\sum\limits_{m}f_m\,{\cal A}_m(L)$
                                                        \\
  Inverse powers
         & $\left[a(L)\right]^{-m}\vphantom{^{\big|}_{\big|}}$
         & ~{---}~
         & ${\cal A}_{-m}(L)=L^m$
                                                        \\
  Multiplication
         & $a^{\mu} a^{\nu}= a^{\mu+\nu}\vphantom{^{\big|}_{\big|}}$
         & ~{---}~
         & ~{---}~                                      \\
  Index derivative
         & $a^{\nu} \ln^{k}a\vphantom{^{\big|}_{\big|}}$
         & ~{---}~
         & $\frac{d^{k}{\cal A}_\nu}{d\nu^{k}}
          =\left[a^{\nu}\ln^{k}(a)\right]_{\rm an}$
                                                        \\
  \hline
\end{tabular}
\label{Tab:char}
\end{table}
%%%%%%%%%%%%%%%%%%%%%%%%%%%%%%%%%%%%%%%%%%%%%%%%%%%%%%%%%%%%%%%%%%%%%%%

\section{FAPT Applications at the NLO and Beyond}
\label{Sect:Appl}

In this section, we concentrate on applications of the presented FAPT
formalism to two QCD processes beyond the leading order of perturbative
QCD.

The first process to be considered is the factorizable part of the
pion's electromagnetic form factor at NLO accuracy in Euclidean space.
This process has been widely discussed in the literature using
various techniques---see e.g., \cite{BPSS04} for a recent comprehensive
analysis and comparison with the available experimental data.
At leading twist, one has the convolution 
($A(z)\otimes{z}B(z) \equiv \int_0^1 dz A(z) B(z)$)
\begin{equation}
  F_{\pi}^{\rm Fact}(Q^2)
=
  \varphi_\pi(x,\mu_{\rm F}^2)\otimes
  T^{\rm NLO}_{\rm H}\left(x,y,Q^2;\mu_{\rm F}^2,\mu_{\rm R}^2\right)
                        \otimes
  \varphi_\pi(y,\mu_{\rm F}^2)\, ,
\label{eq:F-fact}
\end{equation}
%Eq (15)
where the twist-two pion distribution amplitude (DA)
(using $\bar x\equiv 1-x$)
\begin{eqnarray}
 \varphi_\pi(x,\mu^2)
  = 6 x \bar x
     \left[ 1
          + a_2(\mu^2) \, C_2^{3/2}(2 x -1)
          + a_4(\mu^2) \, C_4^{3/2}(2 x -1)
          + \ldots
     \right]
\label{eq:phi024mu0}
\end{eqnarray}
%Eq (16)
contains all non-perturbative information on the pion quark structure
in terms of the Gegenbauer coefficients $a_n$, determined at some
typical hadronic scale $\mu^2 \approx 1$~GeV${}^{2}$
\cite{BMS01,BMS04,BMS05lat}.
Note that the quantity $F_{\pi}^{\rm Fact}(Q^2)$ depends, beyond the
LO, on two scales:
the factorization scale $\mu_{\rm F}$ and the renormalization scale
$\mu_{\rm R}$.

To appreciate the differences among the various analytization schemes,
consider the scaled hard-scattering amplitude entering Eq.\
(\ref{eq:F-fact}) in the so-called ``Naive Analytization'' (Naive-An)
scheme \cite{SSK99,SSK00} in comparison with the ``Maximal
Analytization'' (MA) scheme \cite{BPSS04} (with the renormalization
scale set equal to
$\mu_{\rm R}^{2}=\lambda_{\rm R}Q^2$, $\lambda_{\rm R}$ being a 
numerical parameter):
\begin{eqnarray}
{\it Naive Analytization}~~~~~~~~~~~~~~~~~~~~~\nonumber \\
 \left[Q^2 T_{\rm H}\left(x,y,Q^2;\mu_{\rm F}^2,\lambda_{\rm R} Q^2
                    \right)
 \right]_{\rm Naive-An}
\!\!& = &\!\!
  {\cal A}_{1}^{(2)}(\lambda_{\rm R} Q^2)\,
  t_{\rm H}^{(0)}(x,y)
\nonumber \\
&& \!\!\!\!\!\!\!\!\!\!\!\!\!\!\!\!\!\!\!\!\!\!\!\!\!\!\!\!\!\! +
  \frac{\left[{\cal A}_{1}^{(2)}(\lambda_{\rm R} Q^2)\right]^2}{4\pi}\,
   t_{\rm H}^{(1)}\left(x,y;\lambda_{\rm R},
   \frac{\mu_{\rm F}^2}{Q^2}\right)\\
{\it Maximal Analytization}~~~~~~~~~~~~~~~~~\nonumber \\
  \left[Q^2 T_{\rm H}\left(x,y,Q^2;\mu_{\rm F}^2,\lambda_{\rm R} Q^2
                     \right)
  \right]_{\rm Max-An}
\!\! & = &\!\!
  {\cal A}_{1}^{(2)}(\lambda_{R} Q^2)\,t_{\rm H}^{(0)}(x,y)
\nonumber \\
&& \!\!\!\!\!\!\!\!\!\!\!\!\!\!\!\!\!\!\!\!\!\!\!\!\!\!\!\!\!\! +
  \frac{{\cal A}_{2}^{(2)}(\lambda_{\rm R} Q^2)}{4\pi}\,
  t_{\rm H}^{(1)}\left(x,y;\lambda_{\rm R},
  \frac{\mu_{\rm F}^2}{Q^2}\right)\,.
\label{eq:naiv-vs-max}
\end{eqnarray}
%Eq (17), Eq (18)
In these equations, $t_{\rm H}^{(0)}(x,y)$ and
$t_{\rm H}^{(1)}\left(x,y;\lambda_{\rm R},
\frac{\mu_{\rm F}^2}{Q^2}\right)$
stand, respectively, for the LO and NLO hard-scattering amplitudes,
computed in \cite{MNP99}.

\begin{figure}[th]
 \centerline{\includegraphics[width=0.480\textwidth]{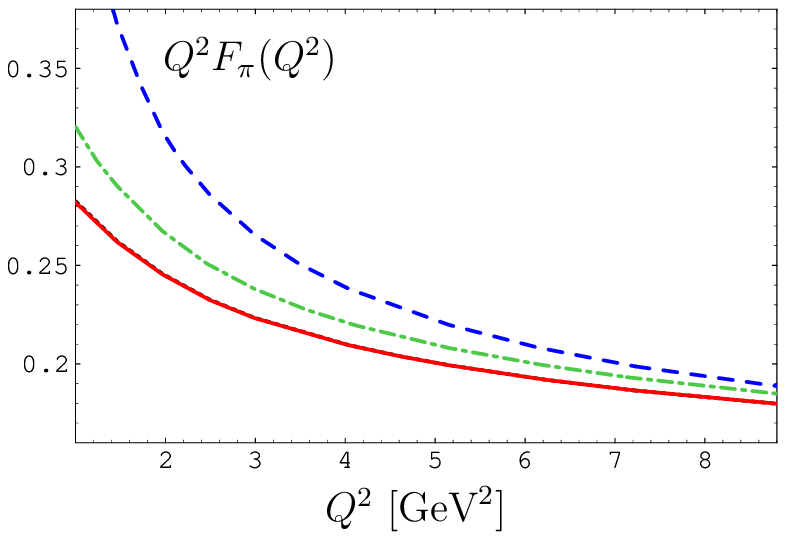}
  %\vspace*{5mm}
 ~~~\includegraphics[width=0.480\textwidth]{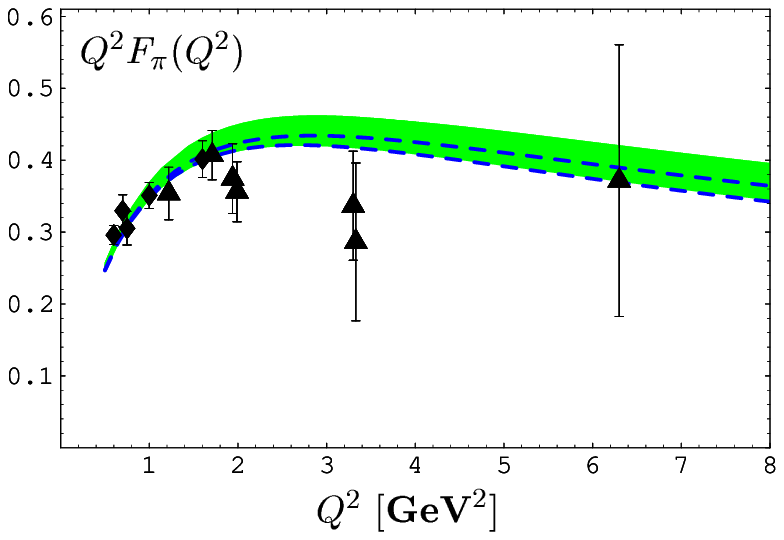}}
 %\vspace{0.0cm}
   \caption[*]{(Left) Results for $Q^2F_{\pi}^{\rm Fact}(Q^2)$ vs.\
    $Q^2$ with $\mu_{\rm R}^2=Q^2$, $\mu_{\rm F}^2=5.76$~GeV$^2$ in
    SPT of QCD (dashed line), using Naive Analytization (dash-dotted
    line), and with Maximal Analytization (solid line).
    (Right) The same quantity (in Max-An) in comparison with
    experimental data (see \protect\cite{BPSS04}).
    The broken lines denote the region accessible to the asymptotic
    pion DA, while the shaded strip marks the region of predictions
    derived with the pion DAs from nonlocal QCD sum rules
    \protect\cite{BMS01} (cf.\ Eq.\ (\ref{eq:phi024mu0})).
\label{fig:pion-FF}}
\end{figure}

The Naive Analytization just replaces the strong coupling and its
powers by their corresponding analytic images.
This procedure is, strictly speaking, incorrect \cite{SSK00} because
$[{\cal A}_{1}(L)]^n\neq [a_s^n(L)]_{\rm An}$,
owing to their distinct spectral representations.
This scheme, whatever its theoretical shortcomings, works
phenomenologically rather well \cite{SSK99,SSK00}.
Its direct improvement in \cite{BPSS04} adopts instead the Maximal
Analytization, which associates to the powers of the running coupling
their own dispersive images, trading this way the usual power series
expansion for a non-power functional expansion, i.e.,
$[a_{s}^{n}(L)]_{\rm Max-An}={\cal A}_{n}(L)$.
The difference between the two analytization schemes becomes apparent
by comparing in the equations above the NLO terms.
This theoretical improvement entails a phenomenological improvement as
well.
From Fig.\ \ref{fig:pion-FF} we see that the crucial advantage of 
the FAPT analysis is that the dependence of the prediction for
$F_{\pi}^{\rm Fact}(Q^2)$ on the perturbative scheme and scale
setting is diminished already at NLO.
Next we will show that applying the KS analytization procedure, the
result will become insensitive also to the variation of the
factorization scale.

%%%%%%%%%%%%%%%%%%%%%%%%%%%%%%FIGURE 2%%%%%%%%%%%%%%%%%%%%%%%%%%%%%%%%%
%%%%%%%%%%%%%%%%%%%%%%%%%%%%%%%%%%%%%%%%%%%%%%%%%%%%%%%%%%%%%%%%%%%%%%%

Consistent with this requirement the analytization of the logarithmic
term
$\ln(Q^2/\mu_{\rm F}^2)=
 \ln (\lambda_\text{R} Q^2/\Lambda^2) -
 \ln (\lambda_\text{R}\mu_{\rm F}^2/\Lambda^2)$
has to be performed as well, so that after some manipulations, 
explained in \cite{BKS05}, we obtain
\begin{eqnarray}
 \left[Q^2 T_\text{H}(x,y,Q^2;\mu_{\rm F}^2,\lambda_{\rm R} Q^2)
 \right]_{\rm KS}^{\rm An}
&  = &
  {\cal A}_{1}^{(2)}(\lambda_{\rm R} Q^2)\, t_{\rm H}^{(0)}(x,y)
\nonumber\\
&&  \!\!\!\!\!\!\!\!\!\!\!\!\!\!\!\!\!\!\!\!\!\!\!\!\!\!\!\!\!\!
    \!\!\!\!\!\!\!\!\!\!\!\!\!\!
+
   \frac{{\cal A}_{2}^{(2)}(\lambda_{\rm R} Q^2)}{4\pi}\,
   t_{\rm H}^{(1)}\left(x,y;\lambda_{\rm R},
  \frac{\mu_{\rm F}^2}{Q^2}\right)
\nonumber\\
&& \!\!\!\!\!\!\!\!\!\!\!\!\!\!\!\!\!\!\!\!\!\!\!\!\!\!\!\!\!\!
    \!\!\!\!\!\!\!\!\!\!\!\!\!\!
    +\ \frac{\Delta_{2}^{(2)}
  \left(\lambda_{\rm R} Q^2\right)}{4\pi}\,
      \left[C_{\rm F}\, t_{\rm H}^{(0)}(x,y)  \,
             \left(6 + 2 \ln(\bar{x}\bar{y})\right)
      \right]\,,
\label{eq:TH-KS-6}
\end{eqnarray}
%Eq (19)
with the deviation from the second line in Eq.\
(\ref{eq:naiv-vs-max}) being encoded in the term
\begin{eqnarray}\label{eq:delta2-2}
 \Delta_{2}^{(2)}\left(Q^2\right)
&\equiv&
   {\cal L}_{2}^{(2)}\left(Q^2\right)
  -{\cal A}_{2}^{(2)}\left(Q^2\right)\,\ln\left[Q^2/\Lambda^2\right]
\end{eqnarray}
%Eq (20)
where
\begin{eqnarray}
 {\cal L}_{2}^{(2)}\left(Q^2\right)
& \!\!\! \equiv \! \!\!\! &
   \left[\left(\alpha_{s}^{(2)}\left(Q^2\right)\right)^2
         \ln\left(\frac{Q^2}{\Lambda^2}\right)
   \right]_\text{KS}^{\rm An}
\!  = \! \frac{4\pi}{b_0}
     \left[\frac{\left(\alpha_{s}^{(2)}\left(Q^2\right)\right)^2}
                 {\alpha_{s}^{(1)}\left(Q^2\right)}
      \right]_\text{KS}^{\rm An}\ .
\label{eq:Log_Alpha_2_KS}
\end{eqnarray}
%Eq (21)
Performing the analytization \cite{BKS05}, we find
\begin{eqnarray}
  {\cal L}_{2}^{(2)}\left(Q^2\right)
   = \frac{4\pi}{b_0}\,
      \left[{\cal A}_{1}^{(2)}\left(Q^2\right)
      + c_1\,\frac{4\pi}{b_0}\,f_{\cal L}\left(Q^2\right)
      \right]\, ,
\label{eq:Log_Alpha_2_BMKS}
\end{eqnarray}
%%Eq (22)
where
\begin{eqnarray}
  f_{\cal L}\left(Q^2\right)
   = \sum_{n\geq0}
      \left[\psi(2)\zeta(-n-1)-\frac{d\zeta(-n-1)}{dn}\right]\,
       \frac{\left[-\ln\left(Q^2/\Lambda^2\right)
             \right]^n}{\Gamma(n+1)}
\label{eq:f_MS}
\end{eqnarray}
%%Eq (23)
and $\zeta(z)$ is the Riemann zeta-function.
One can show (see for details \cite{BKS05})) that calculating
$F_{\pi}^{\rm Fact}(Q^2)$ under the proviso of the KS analytization
provides an expression which is extremely stable against variations
of the factorization scale.
Indeed, varying the factorization scale from 1~GeV${}^2$ to
10~GeV${}^2$, the form factor changes by a mere 1.5 percent and reaches
just the level of about 2.5 percent for a (hypothetical)
factorization scale of 50~GeV${}^{2}$.
The sensitivity on the factorization scale using the Maximal
Analytization is also a mild one, but the corresponding variation is,
in round terms, two times larger.
As regards the scale behavior of the form factor, both analytization
schemes yield almost coincident results.
Hence, $\left[Q^2F_{\pi}^{\rm Fact}(Q^2)\right]_{\rm KS}^{An}$ in Fig.\
\ref{fig:pion-FF} cannot be differentiated from
$\left[Q^2F_{\pi}^{\rm Fact}(Q^2)\right]_{\rm Max-An}$.
In concluding this analysis, using FAPT the dependence on all
perturbative scheme and scale settings, including the factorization
(evolution) scale, is diminished already at the NLO level.

We turn now to the second application, this time in Minkowski
space: the decay of a scalar Higgs boson to a $b\bar{b}$ pair at the
four-loop level of the quantity $R_{\rm S}$ from which one can obtain
the width
$\Gamma(\rm H\to b\bar{b})$ \cite{BMS06}.
In this case, we will encounter no ghost singularities---in contrast
to the Euclidean space.
However, the analytic continuation from the spacelike to the timelike
region will entail so-called ``kinematical'' $\pi^2$ terms, whose
contribution may become with increasing order of the perturbative
expansion as important as the expansion coefficients.

To get a handle on the Higgs-boson decay, we consider the correlator
of two scalar currents
$J^{\rm S}_b=\bar{\Psi}_b\Psi_b$
for bottom quarks with mass $m_b$, coupled to the scalar Higgs boson
with mass $M_{\rm H}$ and where $Q^2 = - q^2$:
\begin{equation}
  \Pi(Q^2)
=
  (4\pi)^2 i\int dx {\rm e}^{iq \cdot x}\langle 0|\;T[\;J^{\rm S}_b(x)
  J^{\rm S}_{b}(0)\,]\;|0\rangle \, .
\label{eq:correlator}
\end{equation}
%Eq (24)
Then,
$
  R_{\rm S}(s)
 =
  \textbf{Im}\, \Pi(-s-i\epsilon)/{(2\pi\,  s)}
$
and one can express the width in terms of $R_{\rm S}$, i.e.,
\begin{eqnarray}
  \Gamma({\rm H} \to b\bar{b})
= \frac{G_{\rm F}}{4\sqrt{2}\pi}M_{\rm H}
  m_{b}^{2}(M_{\rm H}) R_{\rm S}(s = M_{\rm H}^2)
\label{decay_rate_for_b}\, .
\end{eqnarray}
%Eq (25)
One, finally, obtains $R_{\rm S}$ via the analytic continuation of the
Adler function $D$ into Euclidean space by applying on it the linear
operation ${\bf A}_{\rm M}$ (equivalently, the integral transformation
$\hat R$), according to the analytization machinery illustrated in 
Fig.\ \ref{fig:APT-FAPT}
This means that one has to calculate the quantity \cite{ChKS97}
\begin{eqnarray}
  \widetilde{R}_\text{S}(s)
\equiv
  \widetilde{R}_\text{S}(Q^2=s,\mu^2=s)
=
  3m^{2}_{b}(s)\left[ 1 + \sum_{n\geq 1}^{} r_{n}~a_{s}^n(s)
               \right]\, ,
  \label{eq:R-s}
\end{eqnarray}
%Eq (26)
where the expansion coefficients $r_n$ contain characteristic
$\pi^2$ terms originating from the integral transformation $\hat{R}$
of the powers of the logarithms appearing in $\widetilde{D}_{\rm S}$.
The latter is related to $\widetilde{R}_{\rm S}(s,s)$ by means of a
dispersion relation.
Notice that these logarithms have two different sources: one is the
running of $a_s$ in $\widetilde{D}_{\rm S}$, while the other is
related to the evolution of the heavy-quark mass $m_{b}^2(Q^2)$.
As a result, the coefficients $r_n$ in (\ref{eq:R-s}) are connected to
the coefficients $d_n$ in $\widetilde{D}_{\rm S}$ (calculable in 
Euclidean space) and to a combination of the mass anomalous dimension 
$\gamma_i$ and the $\beta$-function coefficients $b_j$, multiplied by 
$\pi^2$ powers \cite{BKM01,Che96,BCK05,ChKS97}.

The running mass in the $l$-loop approximation, $m_{(l)}$, can be
cast in terms of the renormalization-group invariant quantity
$\hat{m}_{(l)}$ to read
\begin{eqnarray}
  m_{(l)}^2(Q^2)
&=& \hat{m}_{(l)}^2
  \left[a_{s}(Q^2)\right]^{\nu_0}
  f_{(l)}(a_s(Q^2))\, ,
\label{eq:m2-hat-run}
\end{eqnarray}
%Eq (27)
where the expansion of $f_{(l)}(x)$ at the three-loop order is given by
\begin{eqnarray}
  f_{(l)}(a_s)
&=&
  1 +  a_s\,\frac{b_1}{2b_0}\left(\frac{\gamma_1}{b_1}
    - \frac{\gamma_0}{b_0}\right)
    +  a_s^2\,\frac{b_1^2}{16\,b_0^2}
            \left[\frac{\gamma_0}{b_0}-\frac{\gamma_1}{b_1}
                 + 2\,\left(\frac{\gamma_0}{b_0}
                 -\frac{\gamma_1}{b_1}\right)^2
\right. \nonumber \\
&& \left.\!\!\!  + \frac{b_0 b_2}{b_1^2}\left(\frac{\gamma_2}{b_2}
                 -\frac{\gamma_0}{b_0}\right)
           \right]
    + O\left(a_s^3\right)\, .
\label{varphi}
\end{eqnarray}
%Eq (28)
We are now ready to consider the analytization of the Adler function
\begin{eqnarray}
  \widetilde{D}_{\rm S}(Q^2;\mu^2)
&=& 
  3\,m_b^2(Q^2)
               \left[1+\sum_{n \geq 1} d_n(Q^2/\mu^2)~a_{s}^n(\mu^2)
               \right]\, .
\label{eq:D-s}
\end{eqnarray}
%Eq (29)
Expanding the running mass in a power series, according to
\begin{eqnarray}
  m_{(l)}^2(Q^2)
&=& 
  \hat{m}_{(l)}^2\, \left(a_{s}(Q^2)\right)^{\nu_0}
  \left[1 + \sum_{m\geq 1}^{\infty} e^{(l)}_m\, 
        \left(a_{s}(Q^2)\right)^m
  \right] \, ,
\label{eq:Z-2loop}
\end{eqnarray}
%Eq (30)
and choosing $\mu^2=Q^2$, we find
\begin{eqnarray}
  \left[3\,\hat{m}_b^2\right]_{(l)}^{-1}\, 
  \widetilde{D}^{(l)}_{\text{S}}(Q^2)
\!\!\!\!\!\! && = 
   \left(a^{(l)}_{s}(Q^2)\right)^{\nu_0}
  +\sum_{n\geq1}^{l}d_n\,
   \left(a^{(l)}_{s}(Q^2)\right)^{n+\nu_0}
\nonumber \\
&& ~~ +\sum_{m\geq1}^{\infty}\Delta^{(l)}_m\,
   \left(a^{(l)}_{s}(Q^2)\right)^{m+\nu_0}   
\label{eq:D-approx}
\end{eqnarray}
%Eqs (31) 
with 
\begin{eqnarray}
  \Delta^{(l)}_m
&=& e_m^{(l)}
  + \sum_{ k\geq1}^{{\rm min}[l,m-1]}d_k\,e_{m-k}^{(l)}\, .
\label{eq:d-tild_1}
\end{eqnarray}
%Eq (32)
Note that we have purportedly separated the mass-evolution
effects (collected in the third term of Eq.\ 
(\ref{eq:D-approx})) from the original series expansion of 
$D$ (truncated at $n=l$), the latter being represented by the 
second term on the RHS of Eq.\ (\ref{eq:D-approx}). 
In practice, for $Q\geq 2$~GeV, i.e., for $\alpha_s\leq0.4$, 
the truncation at $m=l+4$ of the summation  (\ref{eq:Z-2loop}) 
produces a truncation error much smaller than 1 percent.

Finally, we obtain $\widetilde{R}_{\rm S}^{\rm MFAPT}$ from 
the quantity 
$\widetilde{D}_{\rm S}^{(l)}(Q^2)$ 
by applying the analytization operation ${\bf A}_{\rm M}$:
\begin{eqnarray}
  \widetilde{R}_\text{S}^{(l)\text{MFAPT}}
& = &  \textbf{A}_\text{M}[D^{(l)}_{\text{S}}]
\nonumber \\
& = & 
       3\,\hat{m}_{(l)}^2\,
       \left[{\mathfrak a}_{\nu_{0}}^{(l)}
             +\sum_{n\geq1}^{l} d_{n}^{}
              {\mathfrak a}_{n+\nu_{0}}^{(l)}
             +\sum_{m\geq1}^{}\Delta_{m}^{(l)}
              {\mathfrak a}_{m +\nu_{0}}^{(l)}                          
       \right]\,, 
\label{eq:R-MFAPT}
\end{eqnarray}
%Eq (33)
where we have used the short-hand notation 
\begin{eqnarray}
  \left[a_s(s)^{\nu}\right]_\text{an} 
=
  {\mathfrak a}_\nu^{(l)}(s)
\equiv 
  \left(\frac{4}{b_0}\right)^{\nu}
  {\mathfrak A}_{\nu}^{(l)}(s)
\label{eq:Che.Anal.Coupl}  
\end{eqnarray}
%Eq (34)
and $b_0=\frac{11}{3}C_{\rm A} - \frac{4}{3}T_{\rm R}N_f$ with 
$C_{\rm A}=N_c=3$, $T_{\rm R}=\frac{1}{2}$.
The above expression contains, by means of the coefficients 
$\Delta_{n}^{(l)}~(~e^{(l)}_k)$
and the couplings
${\mathfrak a}_{n+\nu_{0}}^{(l)}$, all renormalization-group terms
contributing to this order, while the resummed $\pi^2$ terms are 
integral parts of the \emph{analytic} couplings by construction.

%%%%%%%%%%%%%%%%%%%%%%%%%%%F I G U R E 3%%%%%%%%%%%%%%%%%%%%%%%%%%%%%%%
\begin{figure}[t]
 \centerline{~\includegraphics[width=0.48\textwidth]{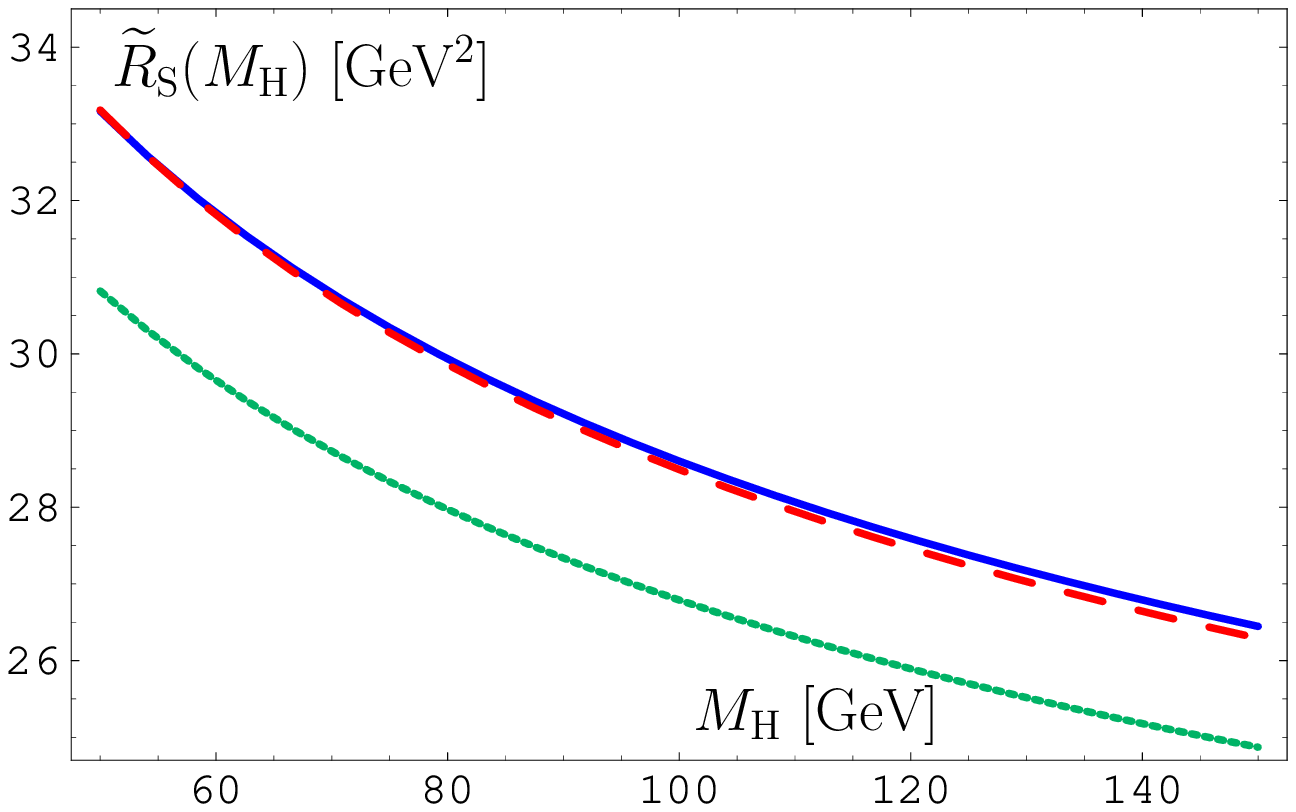}
             ~\includegraphics[width=0.48\textwidth]{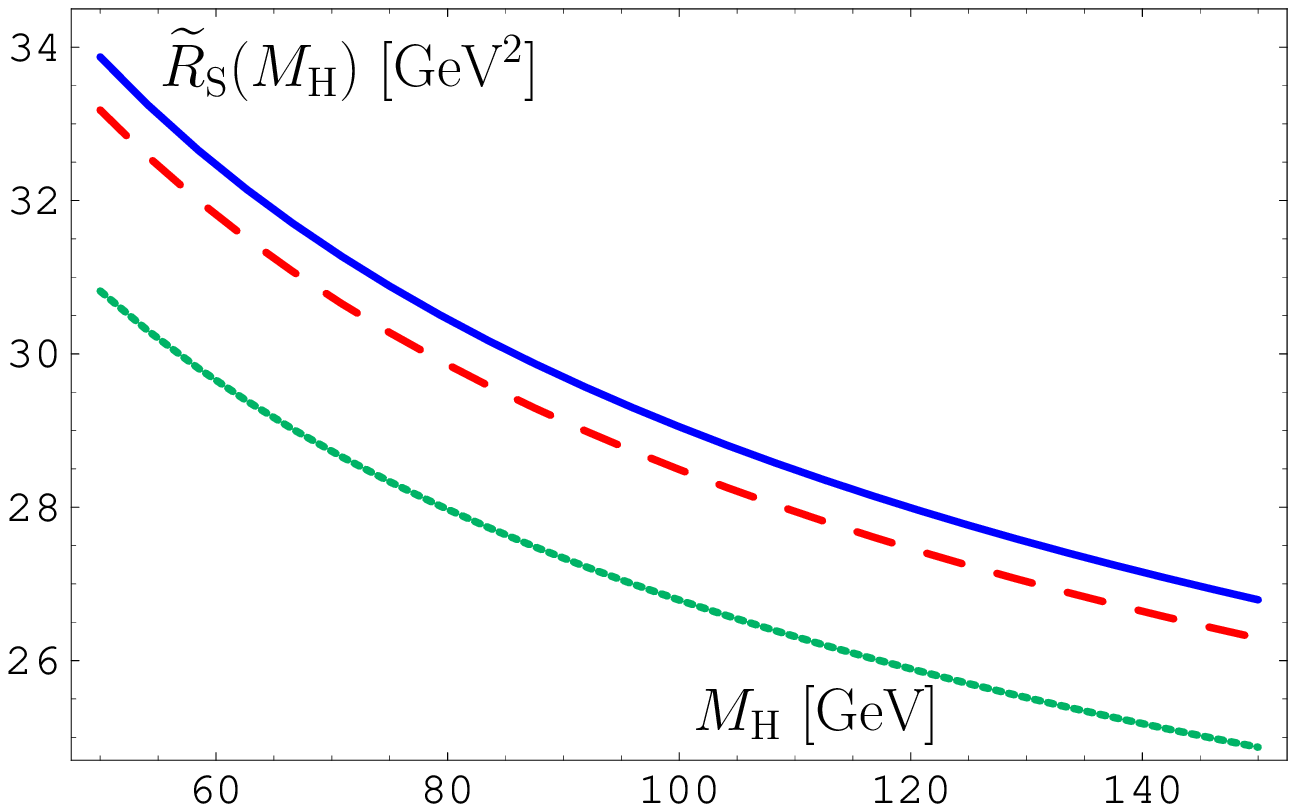}~}
   %\vspace{-0.1cm}
   \caption{Illustration of the calculation of the perturbative series
   of the quantity $\widetilde{R}_{\rm S}(M^2_{\rm H})$ in different
   approaches within the $\overline{\rm MS}$ 
   scheme: 
   Standard perturbative 
   QCD \protect\cite{ChKS97,BCK05} at the loop level $l=4$ (dashed red 
   line with $\Lambda_{N_f=5}=231$~MeV), 
   BKM estimates, by taking into account the 
   $O(\left(a_{s}\right)^{\nu_0}  A_{4}(a_{s}))$-terms, 
   \protect\cite{BKM01}---(dotted green line with 
   $\Lambda_{N_f=5}=111$~MeV), 
   and MFAPT from Eq.\ (\ref{eq:R-MFAPT}) for $N_f=5$ (solid blue 
   line), displayed for two different loop orders: 
   $l=2$ (left panel, $\Lambda_{N_f=5}=263$~MeV) and $l=3$ (right 
   panel with $\Lambda_{N_f=5}=261$~MeV).
   The value of $\Lambda_{N_f=5}$~MeV in all cases corresponds to 
   ${\mathfrak A}_1^{(1)}(s=m_Z^2;N_f=5)=0.120$.}
\label{fig:R_S}
\end{figure}
%%%%%%%%%%%%%%%%%%%%%%%%%%%%%%%%%%%%%%%%%%%%%%%%%%%%%%%%%%%%%%%%%%%%%%%

The results for the quantity 
$\widetilde{R}_{\rm S}(M^2_{\rm H})$, 
calculated within different approaches in the $\overline{\rm MS}$ 
scheme, versus the Higgs mass $M_{H}$ are illustrated in 
Fig.\ \ref{fig:R_S}.
The long-dashed curve in this figure shows the predictions obtained
by Baikov, Chetyrkin, and K\"{u}hn \cite{BCK05} employing standard
perturbative QCD at the $l=4$ loop level of expansion.
The solid curve next to it represents the outcome of the FAPT 
machinery (cf.\ (\ref{eq:R-MFAPT})), including in the second sum all 
evolution effects up to $m=l+4$ and fixing the active flavor number 
to $N_f=5$. 
Bear in mind that the $\pi^2$ terms, induced through the analytic
continuation, are contained in the expansion coefficients 
${\mathfrak a}_{m+\nu_{0}}^{(l)}$.
On the other hand, the contributions of the higher-loop 
renormalization-group dependent terms are accumulated in the 
coefficients $\Delta_{m}^{(l)}$ by means of the parameters 
$\gamma_i$ and $b_j$.  
It is obvious that for this observable the standard perturbative
QCD approach and FAPT yield similar predictions, starting with the 
two-loop running. 
The reason for the slightly larger FAPT prediction lies in the fact
that the coefficients ${\mathfrak a}_{\nu}^{}$ contain the resummed 
contribution of an \textit{infinite} series of $\pi^2$-terms that 
renders them ultimately smaller than the corresponding powers of the 
standard coupling. 
[The interested reader is referred to \cite{BMS06} for further 
details.]
Finally, the lower (green dotted) curve by about 8\% in both panels 
of Fig.\ \ref{fig:R_S} gives the estimate of Broadhurst, Kataev, and 
Maxwell (BKM) \cite{BKM01}, which relies upon the so-called ``naive 
non-Abelianization'' and an optimized power-series expansion that 
makes use of the ``contour integration technique''.\footnote{We thank
A.\ L.\ Kataev for useful remarks pertaining to this figure in 
\cite{BMS06}.} 
[Some more technical remarks can be found in \cite{BMS06}, where the
common elements between this approach and FAPT are worked out.]

\section{Conclusions}
\label{Sect:Con}

The generalized KS analyticity requirement \cite{KS01} has proven 
successful in describing hadronic observables at the partonic level 
for a variety of reactions.
Although this requirement has to be extra postulated, it is the one
that provides a natural extension of the analyticity demand on the
running coupling, proposed by Shirkov and Solovtsov \cite{SS97},
giving us a much broader understanding of analytization.
We have shown that including into the dispersion relations the 
contributions stemming from all terms that affect the spectral density
(even though these terms do not influence the nature of the ghost 
singularities of the standard power-series perturbative expansion),  
makes it possible to treat processes containing two large momentum 
scales.
Such additional scales, like the factorization or the evolution
scale, enter in the form of typical logarithms whose incorporation
into the spectral density naturally amounts to non-integer 
(fractional) powers of the coupling.
This analytization formalism---Fractional Analytic Perturbational 
Theory, developed in \cite{BMS05,BKS05,BMS06} on the theoretical
basis of \cite{KS01}---works equally well in both the spacelike 
region (Euclidean space) as well as the timelike region (Minkowski 
space).
In the first case, the obtained expressions for the hadronic 
observables are singularity-free and turn out to be insensitive to the 
renormalization scheme and scale adopted, while bearing little 
sensitivity to the factorization scale, as well.
In the timelike regime---where ghost singularities are absent---a 
better stability is achieved in terms of expansion coefficients up
to a high loop order that in situ resum all $\pi^2$ terms induced 
by analytic continuation.

In this short exposition we have not been exhaustive.
We note in passing that we have derived closed-form expressions for
the analytic-coupling images at the one-loop level in the spacelike
\cite{BMS05} and in the timelike region, and further approximate 
expressions at the two-loop level \cite{BMS06}, the latter supported 
by exact numerical results \cite{Mag03u}.
What is perhaps more, our approach provides a handle on the computation
of power corrections and their coefficients to different hadronic 
reactions.
We have already obtained leading-order power corrections to the 
pion's electromagnetic form factor and to the cross section of the 
Drell-Yan process into a lepton pair \cite{KS01,Ste03}, which put the 
developed scheme into a larger theoretical and phenomenological 
context.
          
\vskip5mm
\noindent\textbf{Acknowledgments} \\
\vskip-4.5mm \noindent We wish to thank A.P.\ Bakulev and S.V.\ 
Mikhailov for collaboration on some of the issues reported here.
A.I.K. acknowledges financial support from the Pythagoras~I Program 
(Grant 016).
The presented investigations were partially supported by the
Heisenberg-Landau Program (grants 2007 and 2008) and the Deutsche
Forschungsgemeinschaft under contract 436RUS113/881/0.

\end{document}